\documentclass[preprint2]{aastex}
\usepackage{amsfonts}
\usepackage{amssymb}
\usepackage{subfigure}
\usepackage{graphicx,color}

\newcommand{\h}{\ensuremath{\,{\rm h}}}
\newcommand{\psec}{\ensuremath{\, {\rm s}^{-1}}}

\newcommand{\km}{\ensuremath{\,{\rm km}}}

\newcommand{\Mpc}{\ensuremath{\,{\rm Mpc}}}
\newcommand{\K}{\ensuremath{\, {\rm K}}}

\begin{document}

\title{{\LARGE IslandFAST: A Semi-numerical Tool for Simulating the Late Epoch of Reionization}}

\author{{\large Yidong Xu\altaffilmark{1}, Bin Yue\altaffilmark{2}, Xuelei Chen\altaffilmark{1,3,4}\\}}
\altaffiltext{1}{Key Laboratory for Computational Astrophysics, National Astronomical Observatories, Chinese Academy of Sciences, Beijing 100012, China}
\altaffiltext{2}{National Astronomical Observatories, Chinese Academy of Sciences, Beijing 100012, China}
\altaffiltext{3}{University of Chinese Academy of Sciences, Beijing 100049, China}
\altaffiltext{4}{Center for High Energy Physics, Peking University, Beijing 100871, China} 

%\maketitle
\begin{abstract}
We present the algorithm and main results of our semi-numerical simulation, 
{\tt islandFAST}, which is developed from the {\tt 21cmFAST} \citep{2011MNRAS.411..955M}
 and designed for the late stage of reionization.
The {\tt islandFAST} predicts the evolution and size distribution of the large scale under-dense 
neutral regions (neutral islands), and we find that the late Epoch of Reionization (EoR) proceeds very fast, 
showing a characteristic scale of the neutral islands at each redshift.
Using {\tt islandFAST}, we compare the impact of two types of absorption systems, i.e. the large scale 
under-dense neutral islands versus small scale over-dense absorbers, in regulating the reionization process.
The neutral islands dominate the morphology of the ionization field, while the small scale absorbers
 dominate the mean free path of ionizing photons, and also delay and prolong the reionization process.
With our semi-numerical simulation, the evolution of the ionizing background can be derived
self-consistently given a model for the small absorbers.
The hydrogen ionization rate of the 
ionizing background is reduced by an order of magnitude in the presence of dense absorbers.
\end{abstract}


  %--- Methods: analytical}

% Methods: analytical
% intergalactic medium
% Cosmology: theory
% large-scale structure of Universe
% dark ages, reionization, first stars
\maketitle

\section{Introduction}\label{Intro}

The hydrogen gas in the Universe was reionized by the energetic radiation from galaxies and/or  quasars.
Although the details of this process are still highly uncertain and the nature of the ionizing sources 
is poorly understood, some knowledges have been obtained in the past decades and updated recently.
The temperature and polarization data of the cosmic microwave background (CMB) constrain 
the average redshift of reionization to be $z_{\rm reion} \sim 8$ \citep{2016arXiv160503507P}, 
while observations of high redshift quasar (QSO) 
absorption spectra have marked the completion redshift of the hydrogen reionization to be 
$z \approx 6$ (e.g. \citealt{2006ARA&A..44..415F}). 
On the other hand, measurements of the kinetic Sunyaev-Zel'dovich (kSZ)
effect with the South Pole Telescope (SPT) and the Atacama Cosmology Telescope (ACT), in combination
with the Planck data, have given an upper limit on the duration of the reionization, i.e. $\Delta z < 2.8$
\citep{2016arXiv160503507P}. 
Albeit with the above successes, currently measurements of high-redshift galaxy luminosity function 
are limited to the bright end
(e.g. \citealt{2013ApJ...768..196S,2015ApJ...803...34B}), and the constraints on the
ionization state of the IGM from quasar proximity zones observations (e.g. 
\citealt{2011MNRAS.416L..70B,2015MNRAS.452.1105B}) and the Ly$\alpha$ emitting galaxy surveys 
(e.g. \citealt{2012ApJ...744..179S,2014MNRAS.440.3309D}) are quite weak and highly model-dependent.
Various efforts have been made to explore the 21 cm signatures from the neutral hydrogen present
in the IGM during the EoR, and people pin hope on the low frequency radio experiments
such as the Precision Array for Probing the Epoch of Re-ionization (PAPER; 
\citealt{2010AJ....139.1468P,2015ApJ...809...61A}), the Murchison Widefield Array (MWA; 
\citealt{2013PASA...30....7T,2016MNRAS.460.4320E}), the LOw Frequency ARray (LOFAR; 
\citealt{2013A&A...556A...2V}), the Long Wavelength Array (LWA; \citealt{2009IEEEP..97.1421E}), 
as well as the future Hydrogen Epoch of Reionization Array\footnote{http://reionization.org/} 
(HERA; \citealt{2016arXiv160607473D}) and the 
Square Kilometre Array\footnote{http://www.skatelescope.org/} (SKA; \citealt{2013arXiv1311.4288H}).
These 21 cm experiments will greatly push forward the frontiers of our understanding on reionization.

Theoretically, the ``bubble model'' \citep{2004ApJ...613....1F} may be one of the most commonly-accepted 
scenarios for the reionization process. In the inside-out mode of reionization, 
galaxy formation occurs earlier in regions with higher densities, where the IGM was reionized earlier. 
Therefore, the large scale ionization field is closely related to the 
large scale fluctuations of the density field.
One can associate the ionization redshift, or the ionization status of a given position
at a certain redshift, to the local density. This correlation is also confirmed later by numerical simulations 
\citep{2012arXiv1211.2821B}. Based on this idea and the well-established excursion set theory
\citep{1991ApJ...379..440B,1993MNRAS.262..627L},
the bubble model predicts the growth of the ionized regions (``bubbles'') assuming that ionized 
bubbles are spherical and isolated.
The bubble model provides a reasonable description of the growth of HII regions, showing 
good agreement
with numerical simulation results \citep{2007ApJ...654...12Z}. Furthermore, based on its idea, 
approximate treatments of the three dimensional ionization process have been developed, i.e. the 
so called {\it semi-numerical simulation}s (e.g. \citealt{2007ApJ...654...12Z,2007ApJ...669..663M,
2009MNRAS.394..960C, 2011MNRAS.411..955M,2013RAA....13..373Z}).

Strictly speaking, the basic premise of the bubble model is valid only prior to the percolation of HII regions.
Once the ionized bubbles start to contact each other, the assumption that the ionized regions are 
isolated spherical bubbles break down \citep{2014ApJ...781...97X}.
The inaccuracy of applying the bubble model after percolation was also recognized and studied in detail
in \citet{2016MNRAS.457.1813F}.
To generalize the bubble model, the ``island model'' was developed 
in order to better describe the evolution of neutral regions
that are more isolated during the late stage of reionization
\citep{2014ApJ...781...97X}.  
The island model takes into account the presence of an ionizing background that 
should exist in the late EoR, and predicts the distribution and evolution of large scale
neutral regions (``islands'') that are under-dense regions.

In this work we develop a semi-numerical code, {\tt islandFAST}, to realize
the island model in three dimensions, and to simulate the late process of reionization.
Besides the local ionizing sources as modeled in the bubble model and its semi-numerical
counterpart {\tt 21cmFAST} \citep{2011MNRAS.411..955M}, a key ingredient of the island model
is the inclusion of an ionizing background, which is inevitable after percolation.
Therefore, it is also important to incorporate the ionizing background in the semi-numerical 
simulation for the late EoR. Given the flux of the ionizing background and the 
surface area of the region concerned, it is straightforward to compute the ionizations induced by this background. 

The flux of the ionizing background depends on the balance of photon
production and absorption. To generate the ionizing background self-consistently, we take into account the effect of
small scale over-dense absorbers, as well as the large scale under-dense neutral regions -- islands (see \citealt{2012ApJ...747..126A} that first investigated the effect of these both absorbers on the reionization process).

The small scale dense absorbers, which 
are believed to be the main contributor to the IGM opacity \citep{2000ApJ...530....1M,
2005MNRAS.363.1031F,2013ApJ...763..146E}, dominate in
regulating the mean free path of the ionizing photons and hence the intensity of the ionizing background
\citep{2011ApJ...743...82M,2012ApJ...746..125H}. 
They include dense highly non-linear structures, such like interstellar medium (ISM) 
inside galaxies, those mostly ionized, partially self-shielded gas clumps 
in the ionized IGM outside galaxies, usually referred to as LLSs, as well as minihalos or any
other opacity contributors.
The effect of ISM inside galaxies are absorbed in the escape fraction $f_{\rm esc}$ in models.
As the basic idea of the excursion set theory is only valid on large scales, which sets the benchmark
of our island model as well as the {\tt islandFAST}, we adopt an empirical modeling for the 
small scale absorbers in the IGM.

As long as the model of the small scale absorbers is given, the {\tt islandFAST} 
simultaneously generate the ionization field as well as
the ionizing background for the late EoR. 
We further use it as a tool to investigate the roles played by the large scale neutral islands 
as well as the small scale absorbers, in regulating the mean free path
of the ionizing photons and the intensity of the ionizing background, and their effects on the late reionization process.

In the following, we first briefly review the excursion set theory of reionization, i.e.
the bubble model and the island model in section \ref{review_islandmodel}. Then
we describe the algorithm of our semi-numerical simulation, {\tt islandFAST}, in section \ref{islandFAST},
especially the implementation of an ionizing background. The main results of the
simulation are given in section \ref{results}, and we conclude in section \ref{conclusions}.
Throughout this paper, we assume the $\Lambda$CDM model and adopt the following cosmological parameters:
 $\Omega_b = 0.045$, $\Omega_c = 0.225$, $\Omega_\Lambda = 0.73$, 
$H_{\rm 0} = 70\km\psec \Mpc^{-1}$, $\sigma_{\rm 8} = 0.8$ and 
$n_{\rm s} = 0.96$,
but the results are not sensitive to these parameters.

\section{The Excursion Set Theory and the Island Model}\label{review_islandmodel}

The island model is based on the excursion set theory of halo formation. Here we 
briefly review the excursion set approach, especially its application 
to the reionization process, i.e. the bubble model for the early stage of reionization, 
and the island model for the late stage. 
We refer the interested readers to \citet{2007IJMPD..16..763Z} for a detailed
review on the excursion set theory, 
 \citet{2004ApJ...613....1F} and \citet{2005MNRAS.363.1031F} for the bubble model, 
and \citet{2014ApJ...781...97X}  for the island model.

In the excursion set theory, the collapse of a region and formation of halo is determined by 
its average density exceeding a certain threshold (density barrier)  which is a function of redshift and its mass scale. 
In a random density field, the average density around a given position on different smoothing 
mass scales corresponds to a random walk
trajectory in the overdensity-variance plane, and formation of the halo is identified as the first up-crossing of the barrier $\delta_c(M,z)$.  
By computing the {\it first} up-crossing distribution of random walks with respect to the 
 density barrier for halo formation, the excursion set theory recovers the 
 Press-Schechter formula of halo mass function 
at any given redshift, 
and naturally solves the so-called ``cloud-in-cloud'' problem in the original Press-Schechter model.

The excursion set theory can also be applied to the reionization process with the
bubble model. The basic idea of the bubble model is, it asks whether a region has produced 
sufficient number of photons to get itself ionized. The number of ionizing photons produced in the 
region are assumed to be proportional to the total number of baryons in the halos, i.e.  
total mass of the region times the collapse fraction \citep{2004ApJ...613....1F}. 
The ionization condition can be written as 
\begin{equation}\label{Eq.bubbleCriterion}
f_{\rm coll} \ge \xi^{-1},
\end{equation}
where 
\begin{equation}
\xi = f_{\rm esc}\, f_\star\, N_{\rm \gamma/H}\, (1+\bar{n}_{\rm rec})^{-1}
\end{equation}
is the ionizing efficiency parameter,  in which $f_{\rm esc}$, $f_\star$, $N_{\rm \gamma/H}$, 
and $\bar{n}_{\rm rec}$ are the escape fraction, star formation efficiency, 
the number of ionizing photons emitted per H atom in stars, 
and the average number of recombinations per ionized hydrogen atom, respectively. 
Assuming Gaussian density fluctuations, the collapse fraction of a region with
mass scale $M$ at redshift $z$ can be 
written as a function of its mean linear overdensity $\delta_{\rm M}$ \citep{1991ApJ...379..440B,1993MNRAS.262..627L}:
\begin{equation}\label{Eq.fcoll}
f_{\rm coll}(\delta_{\rm M}; M,z) = {\rm erfc} \left[
\frac{\delta_c(z)-\delta_{\rm M}}{\sqrt{2[S_{\rm max}-S(M)]}}\right],
\end{equation}
where $\delta_c(z)$ is the linear critical overdensity for 
halo collapse at redshift $z$,
$S (M) = \sigma^2(M)$ is the variance of the density fluctuations 
smoothed on mass scale $M$, it decreases with increasing $M$, and
$S_{\rm max} = \sigma^2(M_{\rm min})$, in which $M_{\rm min}$ is the minimum 
mass of star-forming halos and is usually taken to be the mass corresponding to $10^4 \K$ viral temperature, 
at which point atomic hydrogen line cooling becomes efficient.

With this collapse fraction, the self-ionization condition can be expressed as   
a random trajectory in the $S-\delta$ space exceeding the barrier on 
the density contrast, i.e. the {\it bubble barrier}: $\delta_{\rm M} > \delta_{\rm B}(M,z)$, where
\begin{equation}\label{Eq.bubbleBarrier}
\delta_{\rm B}(M,z) \equiv \delta_c(z) - \sqrt{2[S_{\rm max} - S(M)]} \, {\rm erfc}^{-1} \left(\xi^{-1} \right).
\end{equation}
Solving for the {\it first}-up-crossing probability distribution of random walks with respect 
to this barrier, $f(S,z)$, the size distribution of ionized bubbles is obtained.

The bubble model gives reasonable description of the reionization process before percolation of the ionized regions 
\citep{2014ApJ...781...97X,2016MNRAS.457.1813F}. After percolation, 
when the ionized regions are connected with each other and the neutral islands are
more isolated, one may use the complementary island model to describe the process.
In the island model, we assume that the ionized regions are connected and consider instead isolated neutral regions.
A region remains neutral if the available ionizing photons is fewer than the required
number to ionize all hydrogen atoms in the region.
A key ingredient of the island model is the inclusion of an ionizing background
which is globally produced during the late EoR, and the 
condition for an island of mass scale $M$ at redshift $z$ 
to keep from being totally ionized is modified accordingly: 
\begin{equation}\label{Eq.IslandCondition}
\xi f_{\rm coll}(\delta_{\rm M}; M,z)+ \frac{\Omega_m}{\Omega_b} 
\frac{N_{\rm back} m_{\rm H} }
{M X_{\rm H} (1+\bar{n}_{\rm rec})} < 1,
\end{equation}
where the second term on the L.H.S. accounts for the contribution from the 
ionizing background, with
$N_{\rm back}$ being the number of consumed background ionizing photons
 and $X_{\rm H}$ being the mass fraction of the baryons in hydrogen.

Rewriting the condition Eq. (\ref{Eq.IslandCondition}) as a
constraint on the density contrast of the region using Eq.~(\ref{Eq.fcoll}),
we derive the {\it island barrier}:
$\delta_{\rm M} < \delta_{\rm I}(M,z)$,
\begin{equation}\label{Eq.islandBarrier}
\delta_{\rm I}(M,z) \equiv \delta_c(z) - \sqrt{2[S_{\rm
max} - S(M)]} \, {\rm erfc}^{-1} \left[K(M,z)\right],
\end{equation}
where
$$K(M,z) = \xi^{-1} \left[1 - N_{\rm back} (1+\bar{n}_{\rm rec})^{-1}
\frac{m_{\rm H}} {M (\Omega_b / \Omega_m) X_{\rm H}}\right].$$
Assuming spherical shape for the island, and that the number of background ionizing photons 
consumed by it at any instant is proportional to its surface area, 
we can derive the total number of background ionizing photons consumed:
\begin{equation}\label{Eq.Nback}
N_{\rm back} = \frac{4\pi}{3}  \left(R_i^3-R_f^3\right) \bar{n}_{\rm H} (1+\bar{n}_{\rm rec}),
\end{equation}
where $\bar{n}_{\rm H}$ is the mean hydrogen number density, and 
$R_i$ and $R_f$ denote the initial and final scale of the island, respectively.
The sphere between the scale $R_i$ and $R_f$ is ionized by the ionizing background, 
so that
\begin{equation}\label{eq.dR}
\Delta R \equiv R_i - R_f = \int_z^{z_{\rm back}} \frac{F(z)}
{\bar{n}_{\rm H} (1+\bar{n}_{\rm rec})} \, \frac{{\rm d}z}{H(z)(1+z)^3},
\end{equation}
where $F(z)$ is the physical number flux of background ionizing photons which is related to the
comoving photon number density by $F(z)=n_\gamma(z)\, (1+z)^3\,c/4$. Note that
the island barrier has a different shape from the bubble barrier because of 
the contribution of the ionizing background photons.

In Eq.~(\ref{eq.dR}), $z_{\rm back}$ is the ``background onset redshift'', below which we assume 
that a spatially homogeneous ionizing background flux is established throughout all of the ionized regions.
Here we assume this happened when regions with average density ($\delta=0$) is ionized. Using the 
bubble model, 
this redshift can be solved from the following equation:
\begin{equation}
\delta_{\rm I}(S=0;z=z_{\rm back}) = 0.
\end{equation}
The solution depends on the parameters of reionization model. For example, 
if we take $\{f_{\rm esc}, f_{\star}, N_{\rm \gamma/H}, \bar{n}_{\rm rec}\}
=\{0.3, 0.1, 4000, 3\}$, then $\xi=30$ and $z_{\rm back}=7.90$. The combination of
 $\{f_{\rm esc}, f_{\star}, N_{\rm \gamma/H}, \bar{n}_{\rm rec}\}
=\{0.2, 0.1, 4000, 3\}$ gives $\xi=20$ and $z_{\rm back}=7.10$, 
while $\{f_{\rm esc}, f_{\star}, N_{\rm \gamma/H}, \bar{n}_{\rm rec}\}
=\{0.2, 0.1, 3000, 3\}$ gives $\xi=15$ and $z_{\rm back}=6.51$.

As all trajectories in the excursion set start from the point $(S,\delta)=(0,0)$, 
and islands are identified by {\it down-crossings} of the island barrier which has
a negative intercept,
and the ``island-in-island'' problem is solved naturally by considering only the 
{\it first}-{\it down}-crossings of the barrier curve.
Solving for the first-{\it down}-crossing 
distribution of random trajectories with the island barrier,
one obtains the mass distribution and the volume fraction of neutral regions
at any given redshift after percolation, or below the background onset redshift.

In addition to the ``island-in-island'' problem, there is a ``bubbles-in-island'' effect
in the island model, as there might also be self-ionized regions inside a large neutral island. 
We shall call the island including bubbles as {\it host island}. 
The bubbles inside neutral islands are identified in the excursion set framework
by considering the trajectories which first 
down-crossed the island barrier $\delta_{\rm I}$ at $S_{\rm I}$, then at a larger $S_{\rm B}$
(smaller scale) up-crossed over the bubble barrier $\delta_{\rm B}$.
The size distribution of bubbles inside an island of scale $S_{\rm I}$ and 
overdensity $\delta_{\rm I}$ is characterized by
the conditional probability distribution 
$f_{\rm B}(S_{\rm B},\delta_{\rm B}|S_{\rm I},\delta_{\rm I})$,
which can be similarly computed with a shifted bubble barrier, i.e.
$\delta_{\rm B}^{\prime} = \delta_{\rm B}(S+S_{\rm I}) - \delta_{\rm I}(S_{\rm I})$,
where $S = S_{\rm B} - S_{\rm I}$.
Also, the average bubbles-in-island fraction can be calculated by integrating over
all possible bubble sizes for a given island.

In order to demarcate the scope of application of both bubble model and island model, 
and to define the bona fide neutral islands, we introduced a percolation threshold $p_c$ in \citet{2014ApJ...781...97X}.
The bubble model is considered reliable before bubble filling factor becomes larger than 
the percolation threshold $p_c$, while the island model can make
accurate predictions only below a certain redshift after when the
island filling factor is below $p_c$. The ionizing background was set up sometime in between, 
after the ionized bubbles percolated but before the islands were all 
isolated.  
The percolation threshold is also applied to the bubbles-in-island fraction.
Only those islands with the bubbles-in-island fraction $q_{\rm B} < p_c$ are qualified 
as a whole neutral island, preventing them from percolated through by the bubbles inside them.
This percolation criterion of $q_{\rm B} < p_c$ acts 
as an additional barrier for finding islands, which combines with the basic island barrier
to define the host islands.
The percolation threshold for a Gaussian random field of $p_c = 0.16$ 
\citep{1993ApJ...413...48K} is used, because the ionization field follows the density field 
\citep{2012arXiv1211.2832B}, which is 
almost Gaussian on large scales \citep{2013arXiv1303.5084P}. More recently, 
a percolation threshold of about 0.1 was derived for the reionization process 
by using the semi-numerical code {\tt 21cmFAST} \citep{2016MNRAS.457.1813F}. However,
the basic algorithm of our semi-numerical simulation {\tt islandFAST} does not dependent on this threshold, 
and the main results shown below are not sensitive to this threshold.
To ease direct comparison with our analytical model predictions, here we set our default value of 
$p_c = 0.16$.

With the combined island barrier taking into account the bubbles-in-island effect, and
using an ionizing background model 
calibrated by the observed ionizing background at redshift $\sim 6$, 
the size distribution of the neutral islands at any given redshift $z$ 
when the neutral fraction is below $p_c$ can be obtained.
At a given instant shortly after the neutral islands become isolated, our model predicts
that the size distribution of the islands has a peak of a few Mpc,  depending on the model parameters.  
As the redshift decreases, the small islands disappear rapidly 
while the large ones shrinks, but the characteristic scale of the islands does not 
change much
if we constrain the bubbles-in-island fraction to be lower than $p_c = 0.16$.
Eventually, all these large scale neutral islands are swamped by ionization, 
only compact neutral regions such as galaxies or minihalos remain.

\section{The islandFAST code}\label{islandFAST}

The {\tt islandFAST} is a semi-numerical code  to reproduce the late stage of the reionization process.
It is developed from the  {\tt 21cmFAST} code \citep{2011MNRAS.411..955M} which is based on the bubble model,
but is extended to treat the late state of reionization by the island model. Compared with the {\tt 21cmFAST}, 
the major differences are 
(i) the {\tt islandFAST} uses a two-step filtering algorithm in generating the ionization field in order to 
take the bubbles-in-island effect into account; 
(ii) the effect of absorption systems is taken into account and a self-consistent treatment for the 
ionizing background is incorporated.

The basic steps of {\tt islandFAST} are as follows.
\begin{enumerate}
\item Create the linear density field with the given power spectrum
and the linear velocity field using the standard Zel'dovich approximation 
\citep{1970A&A.....5...84Z,1985ApJS...57..241E,2005ApJ...634..728S}, just as the first steps in {\tt 21cmFAST}.

\item  Use the {\tt 21cmFAST} algorithm to generate the ionization field at a redshift slightly higher than the $z_{\rm back}$, 
i.e. update the density field for the redshift using the first order perturbation theory, assuming 
the baryons trace the dark matter distribution, and filter the bubble field using the bubble barrier. 
The halo-finding step is bypassed to speed up the computation, 
as we are interested in the large scale distribution of the neutral islands.
We use this ionization field as the initial condition for the following steps.

\item  For each redshift step below, use the excursion set approach to generate the host island field with the 
island barrier Eq.~(\ref{Eq.islandBarrier}).

\item  Start with the host island field for a specific redshift, apply the bubble barrier within each host island, 
and generate the bubbles in islands, then we get the final ionization field for this redshift.  
\end{enumerate}

Unlike the {\tt 21cmFAST}, we skip the final step of assigning the partial ionization fraction to each neutral pixel, 
because the collapse fraction computed with the excursion set theory (i.e. Eq.(\ref{Eq.fcoll})) is only accurate
on large scales and should not be used on each pixel.
When the ionization field is generated for a given redshift, the percolation threshold $p_c$ can be applied 
to select those almost neutral and nearly spherical islands, i.e. using the neutral fraction
threshold of $f_{\rm HI}^c = 1 - p_c$ in quantifying the sizes of the islands with the
spherical average method (SAM, \citealt{2007MNRAS.377.1043M,2007ApJ...654...12Z}), 
we identify the bona-fide neutral islands as defined in the island model.
We may also use lower values of the threshold $f_{\rm HI}^c$, and then
the neutral regions will be attributed to larger and more sponge-like islands.
Comparing between the islands with different values of $f_{\rm HI}^c$, one 
reveals the morphological information of the islands.

The evolution of ionizing background depends on the detailed history of reionization, as it depends both on the
photon production rate and the regulation by various absorption systems which limit the mean free path of the photons. 
Conversely, it also greatly affects how the reionization would proceed.  A self-consistent treatment is essential for correct 
modeling of the reionization process. 
Besides the large scale neutral islands that block the propagation of the ionizing photons,
the most frequently discussed absorbers are Lyman limit systems, which have large enough 
HI column density to keep self-shielded (e.g. 
\citealt{2000ApJ...530....1M,2005MNRAS.363.1031F,2013MNRAS.429.1695B}).
Minihalos could also block ionizing photons and contribute to the IGM opacity
\citep{2005MNRAS.363.1031F}.
However, due to their shallow gravitational potential and 
the complex evaporation process, the contribution
from the minihalos is highly uncertain \citep{2003MNRAS.346..456O,1999ApJ...523...54B,
2004MNRAS.348..753S,2005MNRAS.361..405I,2006MNRAS.366..689C,2012ApJ...747..127Y}.
Observationally, the post-reionization intensity of the ionizing background has been 
constrained by the mean transmitted flux 
in the Ly-$\alpha$ forest (e.g. \citealt{2011MNRAS.412.1926W,2011MNRAS.412.2543C}).

We divide the absorption systems into two categories,
the relatively large neutral islands, and the small scale absorbers which are not
resolved in the simulation. 
We use a semi-empirical prescription for the contribution to the mean free path from small scale absorbers, and  
 take into account the effects of both the large scale islands and the small scale absorption systems simultaneously.

Due to the small scale absorbers as well as the shading of neutral islands, 
a neutral region (island) will only be illuminated by ionizing photons emitted within a distance which is 
roughly the mean free path of the photon. The comoving number density of background ionizing photons 
at redshift $z$ can be modeled as the integration of escaped ionizing photons that are emitted from 
newly collapsed objects and survived to the distances between the sources and the position under consideration:
\begin{eqnarray}\label{Eq.n_gamma}
n_\gamma(z) &=& \int_z\, {\rm d}z', \bar{n}_{\rm H}\, \left|\frac{{\rm d}f_{\rm coll}(z')}{{\rm d}z'}\right|\, f_\star\, N_{\rm \gamma/H}\, f_{\rm esc}\, \nonumber\\
&&\times \exp \left[\,-\, \frac{l(z,z')}{\lambda_{\rm mfp}(z)}\right]\, \left( 1 - f_{\rm HI}^{\rm host}\right)\, 
\end{eqnarray}
where $l(z,z')$ is the physical distance between the source at redshift $z'$ and the redshift $z$ under 
consideration, and $\lambda_{\rm mfp}$ is the physical mean free path of the background 
ionizing photons, $f_{\rm HI}^{\rm host}$ is the neutral fraction of the host island field.
The factor  $( 1 - f_{\rm HI}^{\rm host})$ is because only those ionizing photons located outside 
of the host islands could contribute to the ionizing background.

The treatment of the mean free path in this paper differs slightly from the analytic model of
\citet{2014ApJ...781...97X}, where the mean free path was assumed to be from LLS and computed according to 
the \citet{2000ApJ...530....1M} model.  Now we assume 
\begin{equation}\label{Eq.mfp}
\lambda_{\rm mfp}^{-1}(z) = \lambda_{\rm I}^{-1}(z) + \lambda_{\rm abs}^{-1}(z),
\end{equation}
where $\lambda_{\rm I}$ is the mean free path of ionizing photons due to large scale underdense islands, and 
$\lambda_{\rm abs}$ is the mean free path limited by small scale overdense absorbers including the effects of
Lyman limit systems and minihalos, or other opacity contribution which are not resolved 
in our simulation. While in principle one could also develop a model including the evolution of the small scale absorbers, here
we adopt a more empirical approach. 
\citet{2010ApJ...721.1448S} provided a fitting formula for the evolution of mean free path of ionizing 
photons based on their observed number density of Lyman limit systems up to redshift 6, which reads
\begin{equation}\label{Eq.l_abs}
\lambda_{\rm abs} = 50\, \left[ \frac{1+z}{4.5}\right]^{-4.44} [{\rm p Mpc}].
\end{equation}
We adopt this evolutionary form for the cumulative effect of all kinds
of small scale absorbers assuming the LLSs as the main contributor. We further assume that 
the number density of small scale absorbers 
evolves smoothly near the completion of reionization, so the above fitting formula can be 
extrapolated to the late stage of reionization.

The mean free path of ionizing photons and the resultant ionizing background is incorporated
in the {\tt islandFAST} in an iterative procedure.
We start from a trial value of $\lambda_{\rm I}$ for a redshift a bit lower than $z_{\rm back}$, and applying 
the ionizing background model (Eq.(\ref{Eq.n_gamma})) and the island barrier (Eq.(\ref{Eq.IslandCondition})).
We generate the host island field, then compute the mean free path of ionizing photons directly from the host 
island field. Practically, we cast lines (a total of $10^7$ for each realization) with random starting points located  
in ionized regions and with random directions, and calculate 
the distances from the starting points and the ending points where phase transitions occur. From the distribution
of the distances, we find the mean free path by equaling it to the critical value that a fraction $1/e$ of the 
total distances are larger than it. This is consistent with the definition of the optical depth, in the sense that
only a fraction of $1/e$ of the ionizing photons can survive to a distance of the mean free path.
Using the derived $\lambda_{\rm I}$, we apply the updated ionizing background again to find the updated 
host island field. After several iterations, we achieve the converged intensity of the ionizing background
and the host island field of this redshift. 
Then the bubble barrier is applied within each host island to find ionized bubbles in islands, and 
obtain the ionization field of this snapshot.

Note that the change in the size of a host island is an integration of the changing rate , 
which is proportional to the redshift-dependent ionizing background $n_\gamma (z)$ (Eq.~(\ref{eq.dR})) .
We divide the simulated redshift range into small bins, $\Delta z$, and approximate the $n_\gamma(z)/\bar{n}_{\rm H}$
as a constant between $z$ and $z-\Delta z$. 
We use the converged $\lambda_{\rm I}$ from the previous redshift as the first trial value for the next redshift.
$\Delta z$ is adaptive, and in each step, we make sure that $\Delta z$ is small enough, during which period the 
$\lambda_{\rm I}$ does not grow too much, so that the constant approximation for $n_\gamma(z)/\bar{n}_{\rm H}$
is valid. This is guarranteed by requiring $\lambda_{\rm I}$ to achieve convergence, 
i.e. relative error in $\lambda_{\rm I}$
is smaller than 2\%, within two times of iteration.
Therefore, the {\tt  islandFAST} has to be run downward from the background onset redshift, 
and the ionization field for a redshift of interest can not be obtained without computing the previous redshift steps.

\begin{figure*}[htbp]
\centering{
\includegraphics[scale=0.45]{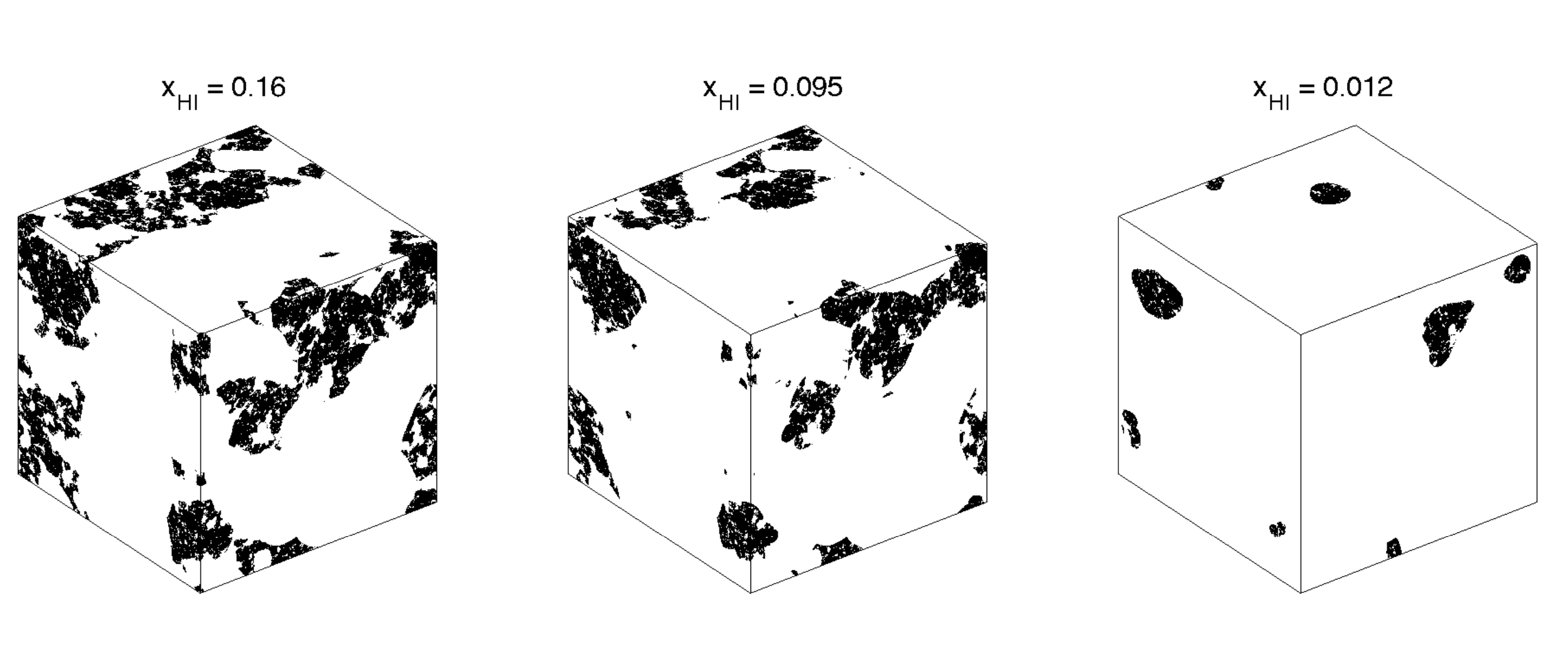}
\caption{The 3-dimensional visualization of the ionization fields from the {\tt islandFAST} with 
a box size of $100 \h^{-1}\Mpc$, $512^3$ resolution, and $\zeta = 20$. 
The neutral islands are shown as black patches, and the ionized regions are left white.
The three boxes have the mean neutral fractions of 
 0.16, 0.095, and 0.012, from left to right respectively.}
\label{Fig.3Dslices_eff20}
}
\end{figure*}

\section{Results}\label{results}

In the default run of {\tt islandFAST}, we take into account the effects of both large scale islands and 
small scale absorbers in regulating the mean free path of ionizing photons. And we set the 
box size of $100 \h^{-1}\Mpc$, and a resolution of $512^3$ for both the dark matter field 
and the ionization field. 
We have made convergence test for the {\tt islandFAST} by running several simulations of different box scales
and resolutions, and find that in terms of the general reionization process and the main results shown
below, convergence is arrived for our default simulation.

\begin{figure}[htb]
\centering{
\includegraphics[scale=0.45]{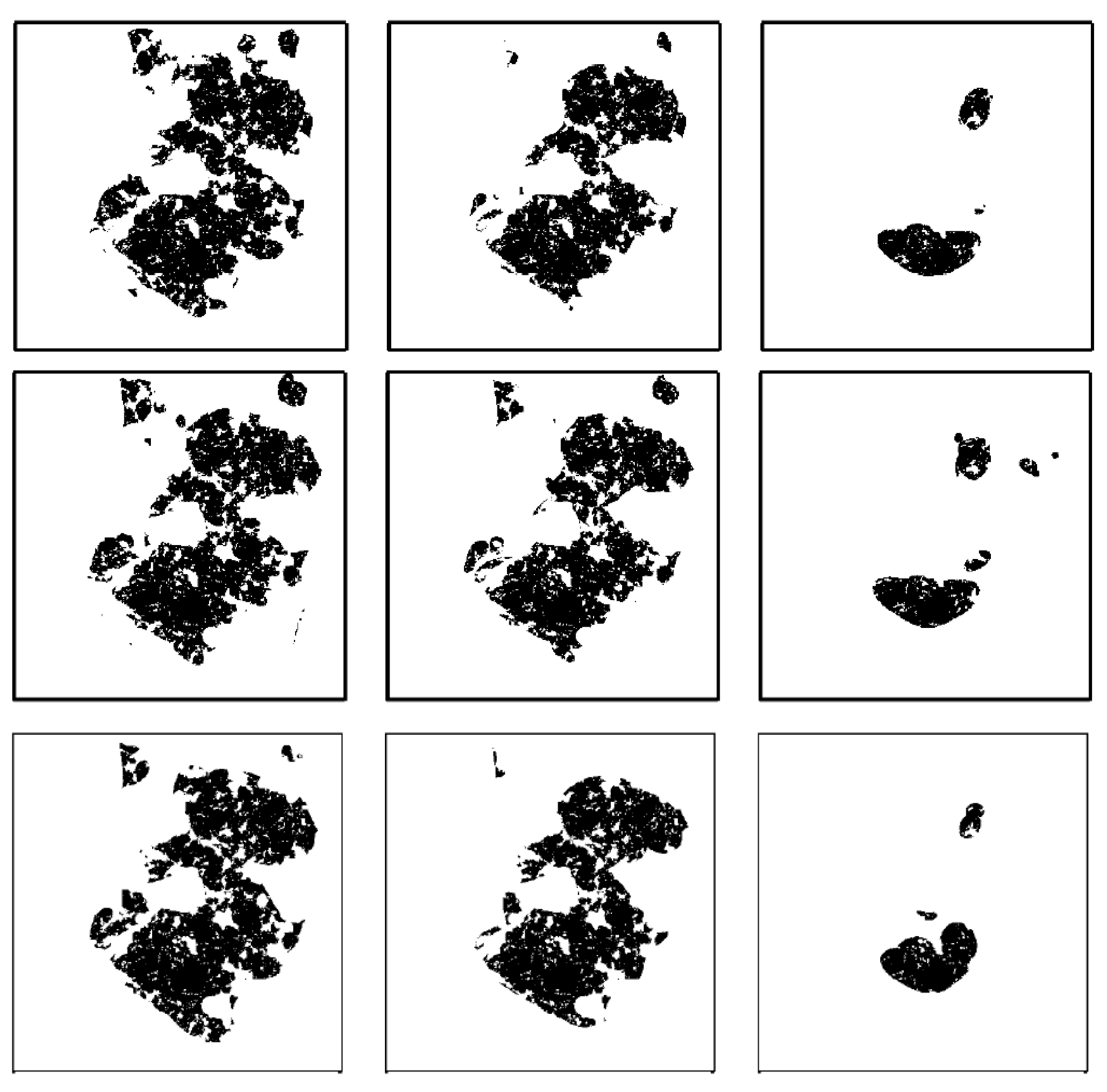}
\caption{Slices of the ionization fields from the {\tt islandFAST}. The top panels are for $\zeta=15$,
middle panels for $\zeta=30$, and the bottom panels are for $\zeta=30$ but without inclusion of 
small scale absorbers. The three columns show the neutral region as the mean neutral fraction decreasing from 
left to right.  The mean neutral fractions are $0.15, 0.093, 0.011$ for the top panels, $0.14, 0.10, 0.013$ for the 
middle panels, and $0.14,0.11, 0.013$ for the bottom panels. }
\label{Fig.slices}
}
\end{figure}

Taking the ionizing efficiency parameter $\zeta = 20$, three example boxes of the ionization 
field at three stages of the late EoR are shown in Fig.~\ref{Fig.3Dslices_eff20}. The black patches
are regions of neutral islands, and the white regions are ionized.
From the left to the right, we see the evolution of the ionization field: the large neutral islands shrink,
while small islands are being ionized and losing their identity as time goes by. The bubbles-in-island effect is obvious 
throughout the late EoR, but as the mean neutral fraction of the Universe decreases, the 
morphology of the ionization field becomes less and less complex, and shape of the islands gradually approaches 
spherical or elliptical.
We also find that the late stage of reionization proceeds quite fast; assuming $\zeta = 20$, 
the mean neutral fraction drops from $\sim 0.16$ to $\sim 0.012$ between $z = 7.0$ and 
$z = 6.425$ in our default run, and the reionization is completed (defined as $x_{\rm HI} < 0.01$) at $z \sim 6.4$.

We compare slices of the simulation box in Fig.~\ref{Fig.slices} for $\zeta=15$ (top panels), $\zeta=30$ (middle panels)
and $\zeta=30$ without small scale absorbers (bottom panels). For each case, we show three slices with decreasing 
mean neutral fraction from left to right. The slices are chosen to show the results of the three different cases 
at about the same mean neutral fraction ($\sim 0.14, 0.10, 0.013$),  though  there are slightly differences due to limitation of simulation step size.

The top panels are from the simulation with $\zeta = 15$, and the middle panels are from the 
simulation with $\zeta = 30$, which is our default run.  Comparing the $\zeta = 15$ case (top panels) with the $\zeta = 30$ (middle panels), 
we find that the morphology of the ionization fields are quite similar at similar mean neutral fraction, 
insensitive to the ionizing efficiency parameter $\zeta$. This can be anticipated because in such models the 
ionization is determined largely by the density field, though it also has some weak dependence on the 
reionization history.

To show the relative impact of large scale islands and small scale absorbers 
on regulating the reionization process, we also run a $\zeta=30$ simulation without the small scale absorbers, 
in which the mean free path of the ionizing background photons is limited only by the neutral islands,
i.e. $\lambda_{\rm mfp} = \lambda_{\rm I}$. The results are  shown in the bottom panels of Fig.~\ref{Fig.slices}
We find that the morphology of the ionization fields are quite similar between the simulations with or 
without small scale absorbers, as long as they are compared at similar neutral fractions, 
implying that the large scale neutral islands are dominant in determining the morphology of the ionization field.
However, the reionization process is much faster in the absence of small absorbers. 
Adopting $\zeta = 30$, the reionization completes at $z_{\rm end} = 7.61$ (when the mean neutral fraction 
$x_{\rm HI} < 0.01$) in the case without small scale absorbers, compared with $z_{\rm end} = 7.07$ 
in the simulation with absorbers. Therefore, the small scale dense absorbers have only moderate effect 
on the morphology of the ionization field at given global neutral fraction, but could delay or prolong the reionization process significantly.

\subsection{Island Size Distribution}

\begin{figure}[t]
\centering{
\includegraphics[scale=0.4]{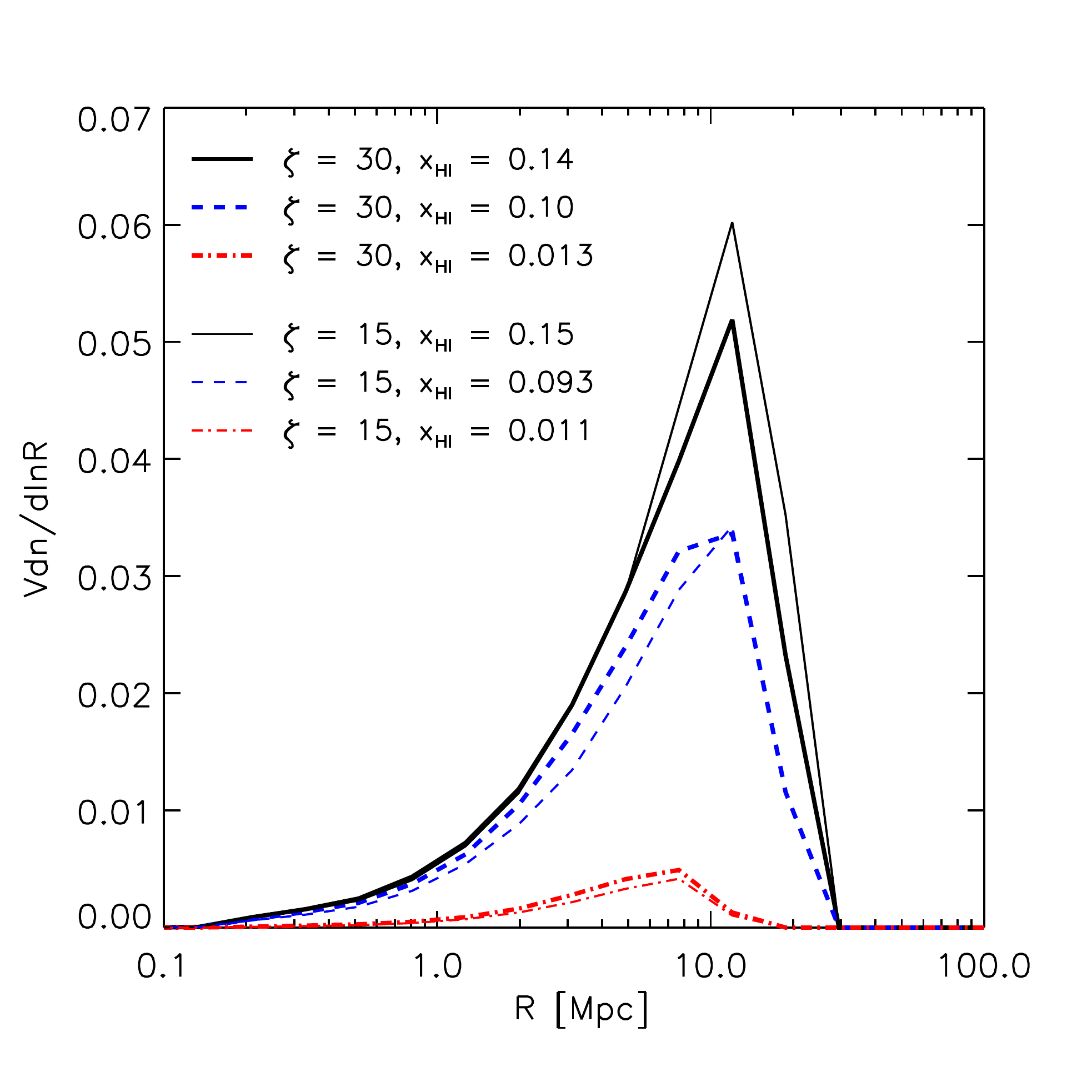}
\caption{The size distribution of neutral islands, using the SAM method with neutral fraction threshold 
$f_{\rm HI}^c = 0.5$. 
The thick lines are from the simulation with $\zeta = 30$,
and the thin lines are from the simulation with $\zeta = 15$.
The solid, dashed, and dot-dashed curves are for the three reionization stages with the mean neutral fractions 
as indicated in the legend.}
\label{Fig.size_SAM_eff30_eff15}
}
\end{figure}

In Fig.~\ref{Fig.size_SAM_eff30_eff15} we show the comoving size distributions of neutral islands for various global
neutral fraction. The neutral islands are selected using the spherical average method (SAM),  
with the critical neutral fraction set to $f_{\rm HI}^c = 0.5$. 
The resulting size distributions of the neutral islands are shown for various global neutral fractions of the Universe with $\zeta=30$ (thick lines)
and $\zeta = 15$ (thin lines). The evolution of the two models characterized by different $\zeta$ values are very similar, which 
is consistent with our impression from the morphology evolution: as the ionization morphology are similar at the same 
neutral fraction, the size distribution should also be similar. In each case,  there is a characteristic scale for the peak of 
neutral island size distribution at each redshift, this scale decreases as the islands are being ionized, but the change 
is very slow. Judging from the simulation box, this is perhaps because the large neutral islands only shrink gradually, 
and as they become smaller they just compensate for the disappearance of the smaller islands.

\begin{figure}[t]
\centering{\includegraphics[scale=0.4]{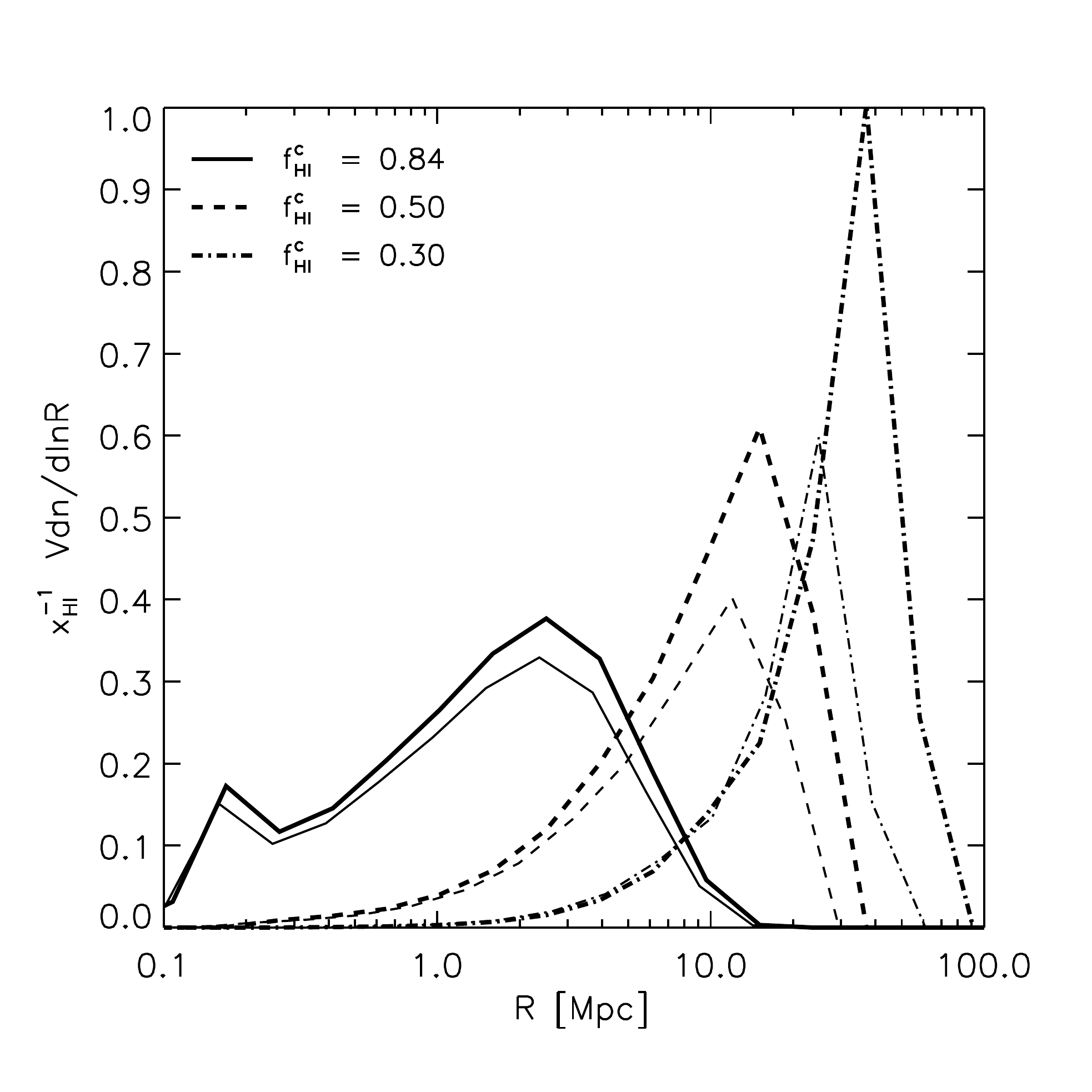}
\caption{The effect of taking different selection thresholds on island size distribution. 
For $\zeta=20$ and $x_{\rm HI}=0.16$, 
the solid, dashed, and dot-dashed curves show the island size distributions with %the selection thresholds of 
$f_{\rm HI}^c = $0.84, 0.5, and 0.3, respectively.
The thick curves are the size distributions of host islands, and the thin curves are those of the net neutral islands.
}
\label{Fig.size_SAM_eff20_thresholds}
}
\end{figure}

However, we must note that the size distribution of the neutral islands depends on the neutral fraction threshold used to
define the islands.  Fig.~\ref{Fig.size_SAM_eff20_thresholds} shows the size distribution of the islands selected by different
neutral fraction thresholds, when the mean neutral fraction of the Universe is fixed at 0.16. 
The thick lines show the size distributions of host islands and the thin
lines show the size distributions of net neutral islands.

\begin{figure}[t]
\centering{\includegraphics[scale=0.4]{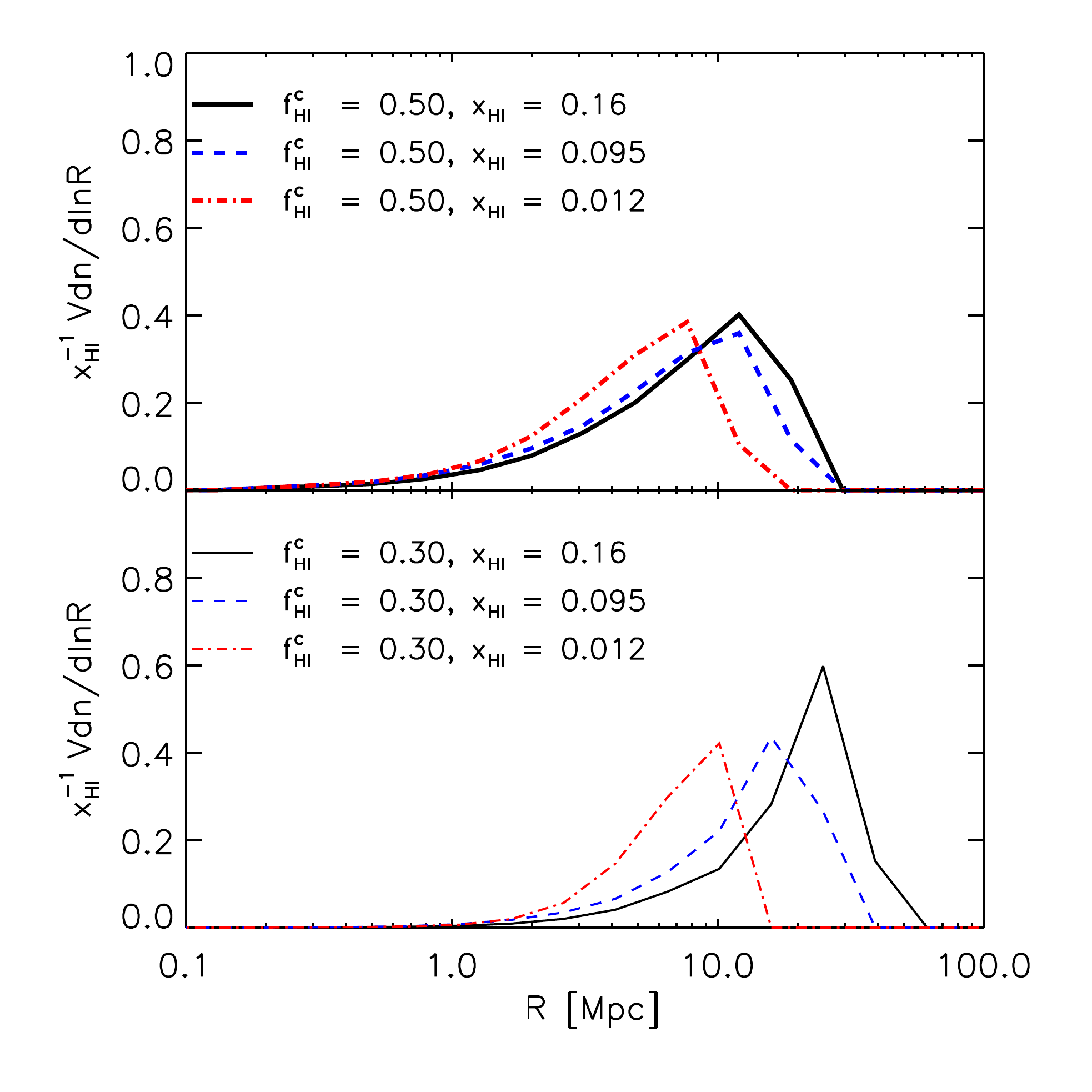}
\caption{The size distribution of neutral islands, obtained with the spherical average method, 
from the simulation with $\zeta = 20$. The {\it upper} panel shows the distribution 
derived with the selection threshold of $f_{\rm HI}^c = $ 0.5, and the {\it lower} panel shows 
the distribution with $f_{\rm HI}^c = $ 0.3.
The solid, dashed, and dot-dashed curves are for the three reionization stages with the mean neutral fractions 
as indicated in the legend.}
\label{Fig.size_SAM_eff20_evol}
}
\end{figure}

\begin{figure*}[t]
\centering{\includegraphics[scale=0.4]{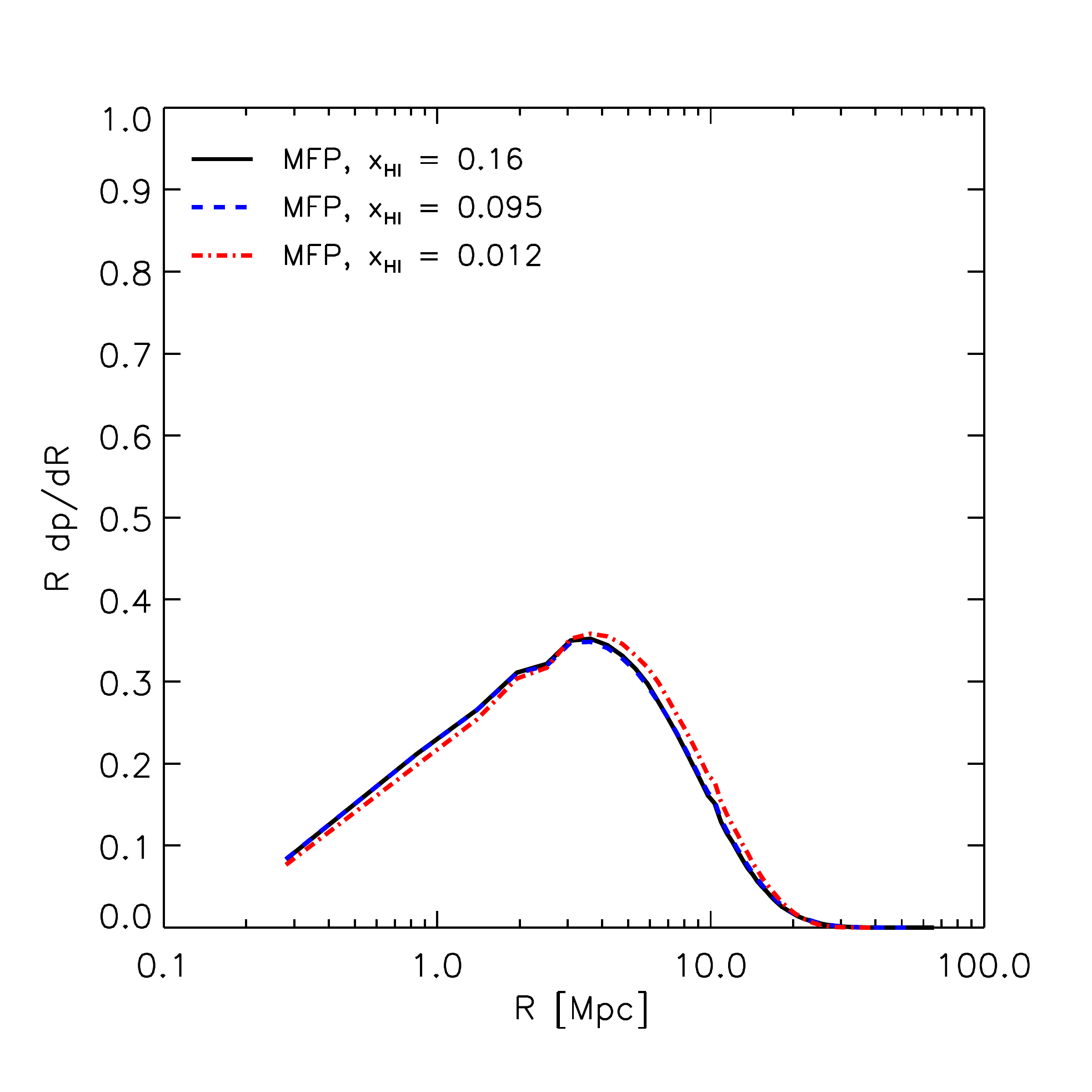}
\includegraphics[scale=0.4]{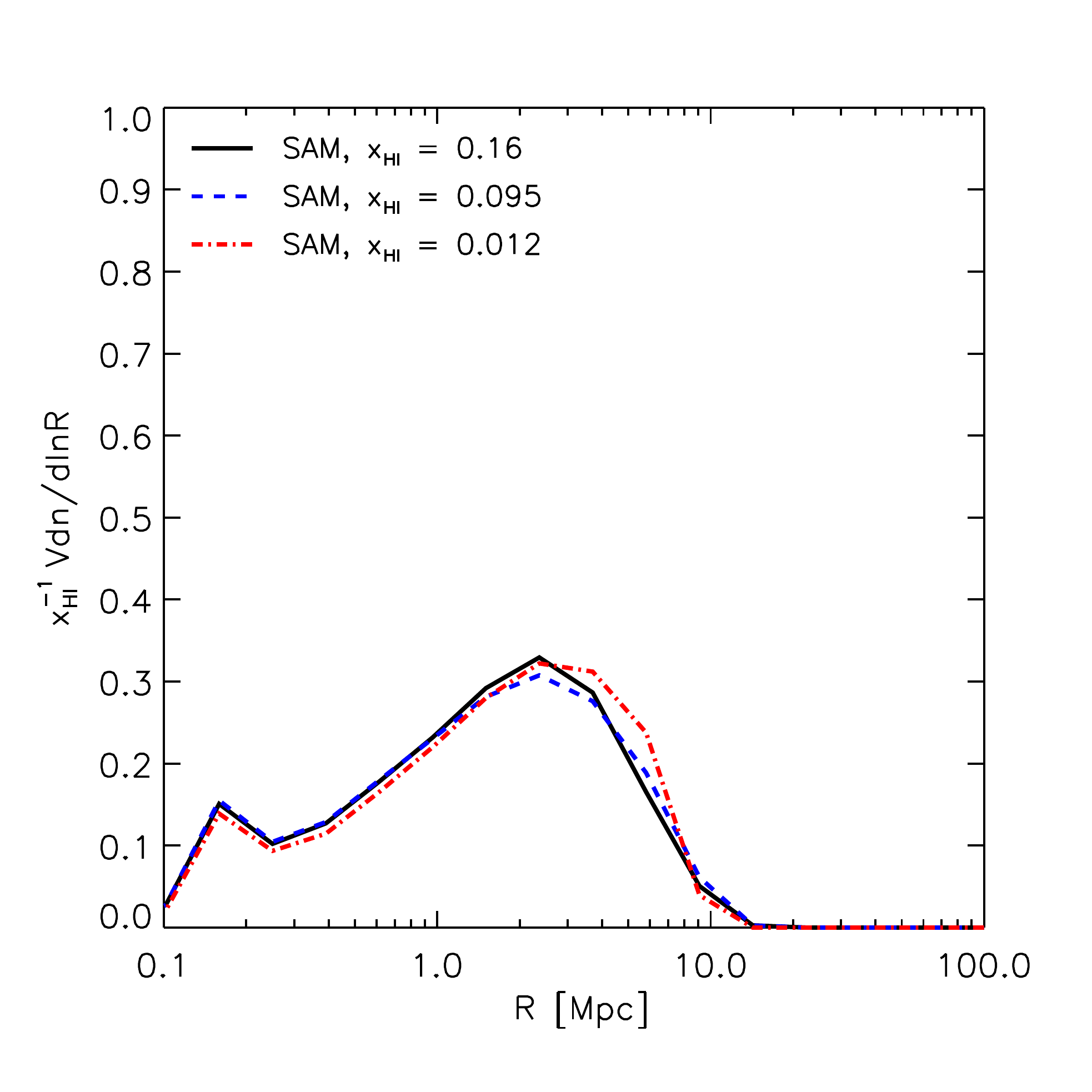}
\caption{The size distributions of neutral islands obtained with the mean free path algorithm ({\it left panel}),
and those obtained with the spherical average method using $f_{\rm HI}^c = 0.84$ ({\it right panel}). 
The solid, dashed, and dot-dashed curves are for the three reionization stages with the mean neutral fractions 
as indicated in the legend, and the ionizing efficiency parameter adopted is $\zeta = 20$.}
\label{Fig.size_MFP_SAM_eff20}
}
\end{figure*}

With lower selection threshold $f_{\rm HI}^c$ value, the evolution of island size is more apparent. The upper and lower panels of Fig.~\ref{Fig.size_SAM_eff20_evol} show the evolution of the size distribution
derived with $f_{\rm HI}^c = 0.5$ and $f_{\rm HI}^c = 0.3$ respectively. 
While the peak size of the neutral islands remains nearly constant when $f_{\rm HI}^c = 0.84$
is used (shown in the right panel of Fig.~\ref{Fig.size_MFP_SAM_eff20}), we find now the peak size does shrink if $f_{\rm HI}^c = 0.3$ is used when selecting the islands.

It is also seen from Fig.~\ref{Fig.size_SAM_eff20_evol} that the difference in the size distributions between
the different selection thresholds is more significant at higher mean neutral fractions, or at earlier time.
This indicates that the shape of the islands is more complex at earlier stage of reionization. 
As the mean neutral fraction decreases, the shape of the islands becomes less and less complex, and
the size of an island becomes less dependent on the selection threshold in the SAM.
The size dependence on the selection threshold of the neutral fraction also implies that in the future
21 cm observations, the higher sensitivity of a radio array could result in a larger typical size 
of the neutral islands, and more evident evolution in the size of neutral patches.

There are various ways of quantifying the size of the ionized bubbles or the neutral islands. 
\citet{2016MNRAS.461.3361L} argued that compared with the spherical averaged result (SAM),
the mean free path (MFP) probability distribution function (PDF) 
is a more physical description of the bubble size. Here we also try the MFP 
to derive the size distribution of the neutral islands, shown in
the left panel of Fig.~\ref{Fig.size_MFP_SAM_eff20} for three stages of the late EoR.
Using the mean free path description, we find an almost constant characteristic scale for the neutral islands
throughout the late EoR. For comparison, we also plot in the right panel of Fig.~\ref{Fig.size_MFP_SAM_eff20}
the SAM size distribution derived with $f_{\rm HI}^c = 1 - p_c = 0.84$, which shows the size distribution
of those almost completely neutral islands as defined in the island model. 
Interestingly, we find an almost non-varying characteristic scale for the neutral islands in this case.
This is consistent with the analytical prediction by the island model.
The peak scale of the MFP PDF is about $4\Mpc$, a bit larger than the typical scale of $\sim 3\Mpc$ 
derived by the SAM with $f_{\rm HI}^c = 0.84$, which is consistent with the expectation
in \citet{2016MNRAS.461.3361L}.

However, this invariance of island size is in conflict with the intuition of shrinking islands 
as seen from the slice maps. Because of the complex shape of the islands and the bubbles-in-islands effect,
the MFP description of the islands underestimate the size of the islands, and tends to predict the size of
those almost neutral patches as the SAM with $f_{\rm HI}^c = 0.84$. This implies that although the 
MFP is a good at representing the ionized bubble sizes and the mean free path of ionzing photons 
during the early stage of reionization, it may not be very useful for characterizing the neutral island 
sizes at the late stage of EOR.

\subsection{Ionizing Background}
%The evolution of the ionizing background: 

\begin{figure}[t]
\centering{\includegraphics[scale=0.4]{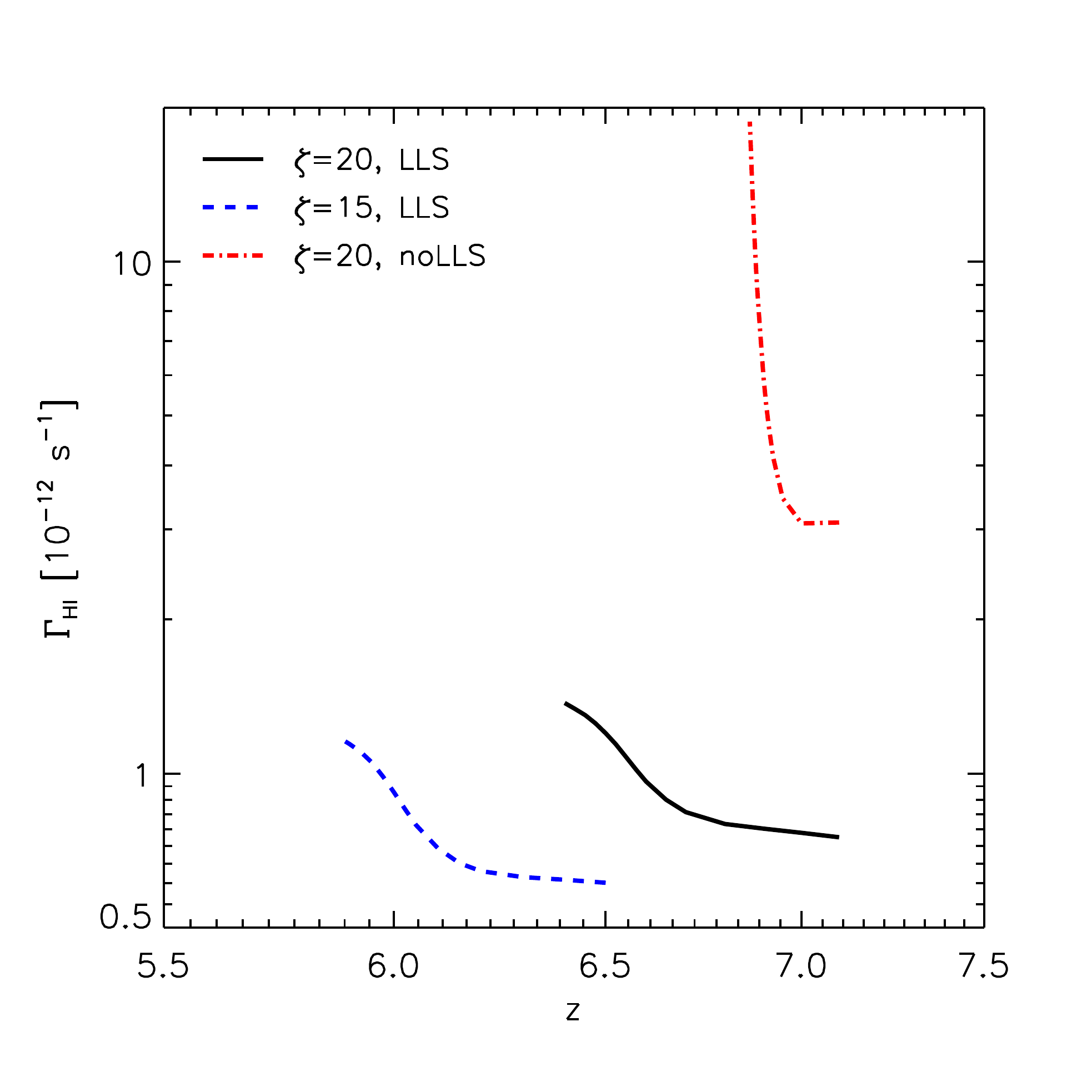}
\caption{The evolution of the ionizing background. The solid and dashed curves show 
the HI photoionization rate $\Gamma_{\rm HI}$ as a function of redshift  
taking into account the small scale absorbers. The solid curve is from the simulations
with $\zeta = 20$, and the dashed curve is from the one with $\zeta = 15$.
The dot-dashed curve shows the evolution of $\Gamma_{\rm HI}$ from the simulation 
without small scale absorbers, and $\zeta = 20$.}
\label{Fig.Gamma_HI}
}
\end{figure}

While generating the ionization field, the {\tt islandFAST} simultaneously predicts the evolution of the
ionizing background over the redshift range simulated. 
The solid and dashed curves in Fig.~\ref{Fig.Gamma_HI} show the HI photoionization rate,
$\Gamma_{\rm HI}$, as a function
of redshift predicted by the simulation with $\zeta = 20$ and $\zeta = 15$ respectively. 
The curves display rapid increase below the background onset redshift, indicating quick growth
of the intensity of the ionizing background during the late EoR. After that, the growth of 
the ionizing background slows down as the reionization approaching the completion.
We find that the intensity of the ionizing background and the timing of its rapid growth depends
significantly on the adopted ionizing efficiency parameter $\zeta$. A higher ionizing efficiency 
would result in a much higher intensity and earlier growth of the ionizing background.

To show the effect of small scale dense absorbers in regulating the ionizing background, in Fig.~\ref{Fig.Gamma_HI}
we also plot with the dot-dashed line the evolution of $\Gamma_{\rm HI}$ predicted by the simulation without small scale
absorbers for $\zeta = 20$. 
The intensity of the ionizing background is boosted by an order of magnitude in the absence of small absorbers,
and the growth of the ionizing background becomes much faster, which results in the rapid
completion of the reionization process. Therefore, we conclude that the small scale absorbers
have played a dominant rule in regulating the level of the ionizing background, and they delay
and prolong the reionization process significantly.

\begin{figure}[t]
\centering{\includegraphics[scale=0.4]{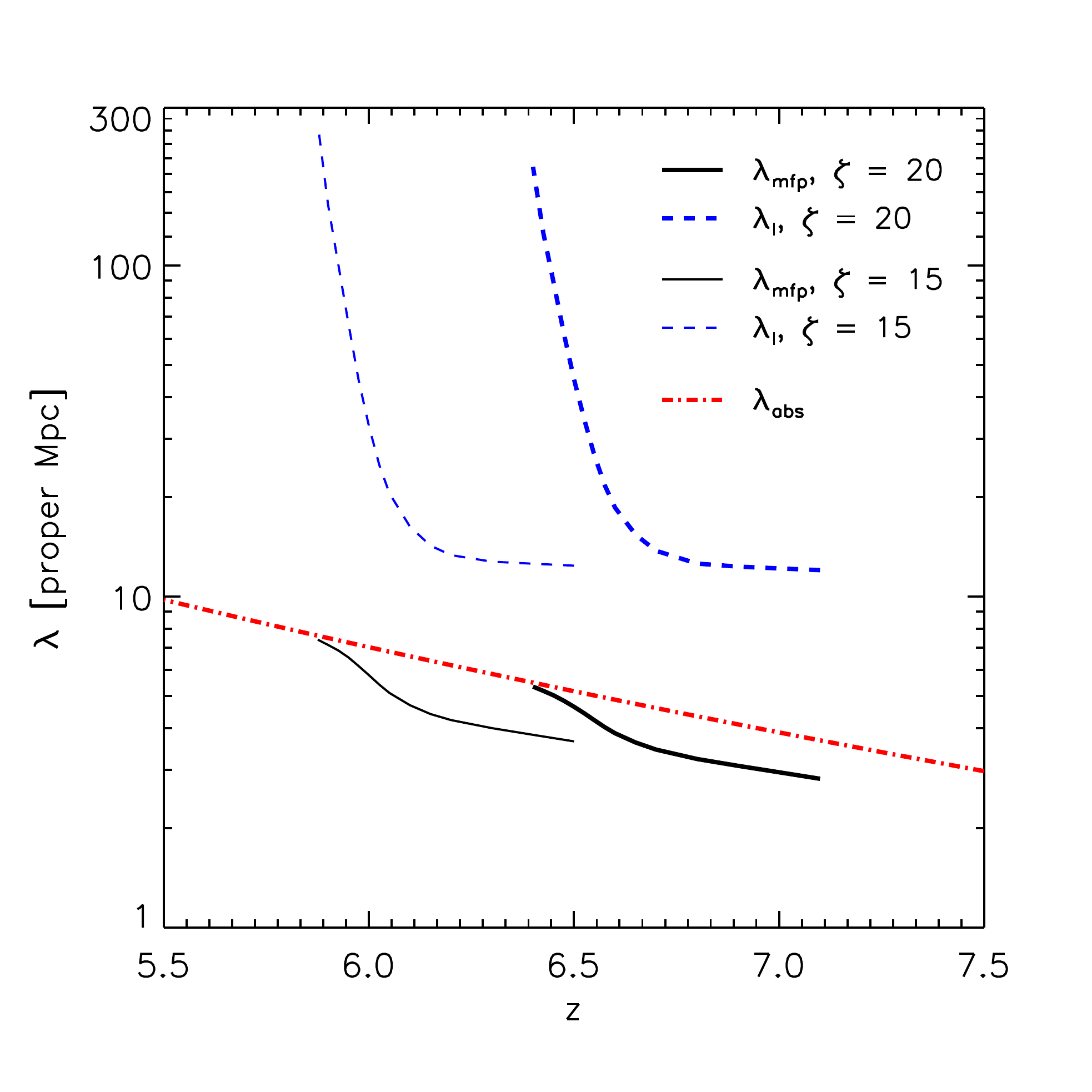}
\caption{The evolution of the mean free path of ionizing photons ({\it solid lines}).
The dashed lines show the mean free path due to neutral islands, while
the dot-dashed line indicates the mean free path due to small scale absorbers 
only, extrapolated from the fitting formula by \citet{2010ApJ...721.1448S}. 
The thick lines are from the simulation with $\zeta = 20$,
and the thin lines are from the simulation with $\zeta = 15$.}
\label{Fig.mfp}
}
\end{figure}

The solid lines in Fig.~\ref{Fig.mfp} show the evolution of the mean free path of the background ionizing 
photons derived from the {\tt islandFAST}, with the thick line from the simulation with $\zeta = 20$
and the thin line from the simulation with $\zeta = 15$. The evolution of the mean free path
shows similar trends as the growth in the intensity of the ionzing background, and
the timing of the growth is also sensitive to the ionzation efficiency.
To reveal the relative importance of the underdense islands and the overdense absorbers in 
limiting the mean free path of the ionzing photons, we plot separately the $\lambda_{\rm I}$
and $\lambda_{\rm abs}$ with the dashed and dot-dashes lines respectively.
We find that the mean free path of the ionzing photons due to islands is always much larger than 
that limited by the small scale absorbers, showing the effect of the latter to be dominant. 
The shading effect of the large scale islands 
reduce the mean free path moderately during the EoR, and as we approach
the end of reionization, the islands are gradually eroded by the ionzing background, 
and the effective mean free path of the ionzing photons approaches the value limited by
the small scale absorbers. Therefore, we confirmed the previous understanding that 
the small scale dense absorbers (probably LLSs) are the main contributor to the IGM opacity
\citep{2005ApJ...624..491I,2013ApJ...763..146E}. 

We note that if $\zeta = 15$ is adopted, for which the reionization completes at $z \approx 5.9$,
the predicted level of the ionizing background is $\Gamma_{\rm HI} \sim 1 \times 10^{-12} \psec$
at redshift 6. This is higher than the observational constrains of 
$\Gamma_{\rm HI} = 0.18^{+0.18}_{-0.09}\times 10^{-12} \psec$ by \citet{2011MNRAS.412.1926W}, 
or $\log (\Gamma_{\rm HI}) = -12.84 \pm 0.18$ by \citet{2011MNRAS.412.2543C}.
This discrepancy may be caused by our brute extrapolation of the mean free path of ionizing photons,
constrained by the number density of LLSs after reionization,
up to the EoR, or by the uncertainty in the quasar modeling when deriving the
observational constraints from the quasar proximity effect, or by some other reasons.
Also, we have adopted a uniform distribution of the ionizing background with the averaged intensity. 
However, the ionizing background should fluctuate significantly at the end of reionization due to the clustering
of the ionizing sources \citep{2015MNRAS.453.2943C} and that of the neutral islands.
A more sophisticated modeling for the ionizing background may be incorporated in line with
\citet{2016arXiv160704218O}.
We defer a more through and quantitative investigation on the ionizing background level to future works.

\section{Conclusions}\label{conclusions}

In this paper, we present the algorithm and some simulation results from a semi-numerical reionization 
simulation code, {\tt islandFAST}, which is designed for the late stage of reionization, 
after the percolation of ionized bubbles. It is an extension of the semi-numerical reionization simulation code
based on the bubble model. The simulation incorporates the effect of ionizing background photons on the 
neutral islands, i.e. underdense regions which are ionized later. It predicts the evolution of the ionization field, 
showing the prevalence of the bubbles-in-island effect.
 
As expected, in the simulation the large islands shrink with time and the small ones are swamped 
by the ionizing photons as reionization processes.
Using either the spherical average method with a high neutral fraction threshold 
(e.g. $f_{\rm HI}^c = 0.84$) or the mean free path PDF, we derive the 
the size distribution of the neutral islands. An interesting result is that these distributions exhibit 
a relatively robust characteristic scale of a few Mpcs throughout the late EoR. 
When the spherical average method is used and 
a lower threshold of neutral fraction is used for island selection, 
 the islands have larger sizes and the size evolves more evidently.

The {\tt islandFAST} generates the intensity of the ionizing background as well as the mean free path of
the ionizing photons simultaneously with the ionization field, 
as long as a reasonable model for the small scale absorbers is provided. 
Therefore, it can be used to investigate the roles played by both the large scale underdense islands and 
small scale overdense absorbers in modulating the ionizing background. 
Neglecting the small absorbers, we provide a self-consistent model for the evolution
of the ionizing background regulated by only the shading effect of large scale islands. 
Taking also the small absorbers into account, the {\tt islandFAST} serves as a tool to model 
the relative contribution of islands and small absorbers to the IGM opacity. 
We find that while the large scale islands dominate the morphology of the ionization field,
it is the small scale absorbers that dominate the opacity of the IGM, and play a 
major rule in limiting the mean free path of the ionizing photons and determining the
intensity of the ionizing background. They also delay and prolong the reionization process.
However, there is still some quantitative discrepancy in the model prediction with current observation results, so at present
the conclusions regarding to the ionizing background should be taken as qualitative.

\acknowledgments
We thank Marcelo Alvarez for providing the spherical average code. YX is supported by 
the NSFC grant 11303034, and the Young Researcher Grant of 
National Astronomical Observatories, Chinese Academy of Sciences. 
BY is supported by the CAS Pioneer Hundred Talents (Young Talents) Program. 
XC is supported by the MoST 863 grant  2012AA121701, 
the NSFC key project grant 11633004 and grant 11373030, the Chinese Academy of Sciences
Frontier Science Key Project QYZDJ-SSW-SLH017 and the Chinese Academy of Sciences
Strategic Priority Research Program XDB09020301.

\bibliography{references}
\bibliographystyle{hapj}

\end{document}